\documentclass{article}
\usepackage[accepted]{icml2022}

\usepackage[utf8]{inputenc} 
\usepackage[T1]{fontenc}    
\usepackage{url}            
\usepackage{booktabs}       
\usepackage{amsfonts,amsmath,bm}       
\usepackage{nicefrac}       
\usepackage{microtype}      
\usepackage{color}
\usepackage[ruled]{algorithm2e} 
\SetKwInOut{KwIn}{Initialize}
\usepackage[colorlinks,
    linkcolor=blue,      
    anchorcolor=blue,  
    citecolor=blue,
]{hyperref}

\usepackage{multirow}
\usepackage{array}
\usepackage{tabularx}
\usepackage{caption}
\usepackage{subfigure}
\usepackage{graphicx}
\usepackage{mdwlist}
\usepackage{epsfig}
\usepackage{epstopdf}
\usepackage{parskip}

\begin{document}

\twocolumn[
\icmltitle{Temporal-Spatial Entropy Balancing for Causal Continuous \\Treatment-Effect Estimation}

\icmlsetsymbol{equal}{*}

\begin{icmlauthorlist}
\icmlauthor{Tao Hu}{yyy}
\icmlauthor{Honglong Zhang}{yyy}
\icmlauthor{Fan Zeng}{yyy}
\icmlauthor{Min Du}{yyy}
\icmlauthor{XiangKun Du}{yyy}
\\
\icmlauthor{Yue Zheng}{yyy}
\icmlauthor{Quanqi Li}{yyy}
\icmlauthor{Mengran Zhang}{yyy}
\icmlauthor{Dan Yang}{yyy}
\icmlauthor{Jihao Wu}{yyy}
\end{icmlauthorlist}

\icmlaffiliation{yyy}{Huolala.cn, Inc}

\icmlcorrespondingauthor{Min Du}{dumin.du@huolala.cn}

\icmlkeywords{Causal Inference, Causal Effect Estimation, Entropy Balancing Method, Continuous Entropy Balancing Method}
\vskip 0.3in
]

\printAffiliationsAndNotice{}

\begin{abstract}
In the field of intracity freight transportation, changes in order volume are significantly influenced by temporal and spatial factors. When building subsidy and pricing strategies, predicting the causal effects of these strategies on order volume is crucial. In the process of calculating causal effects, confounding variables can have an impact. Traditional methods to control confounding variables handle data from a holistic perspective, which cannot ensure the precision of causal effects in specific temporal and spatial dimensions. However, temporal and spatial dimensions are extremely critical in the logistics field, and this limitation may directly affect the precision of subsidy and pricing strategies. To address these issues, this study proposes a technique based on flexible temporal-spatial grid partitioning. Furthermore, based on the flexible grid partitioning technique, we further propose a continuous entropy balancing method in the temporal-spatial domain, which named \textbf{TS-EBCT} (\textbf{T}emporal-Spatial\textbf{E}ntropy \textbf{B}alancing for Causal \textbf{C}ontinue \textbf{T}reatments). The method proposed in this paper has been tested on two simulation datasets and two real datasets, all of which have achieved excellent performance. In fact, after applying the TS-EBCT method to the intracity freight transportation field, the prediction accuracy of the causal effect has been significantly improved. It brings good business benefits to the company's subsidy and pricing strategies.
\end{abstract}
\section{Introduction}
In the marketing environment of the logistics industry, it is crucial to describe the impact of subsidy or pricing on order generation using causal inference techniques. Uplift model has been widely used in predicting causal effects. However, during the use of the uplift model, controlling confounding variables is essential to enhance the accuracy of the model in predicting causal effects. The current popular practice is to use the Inverse Propensity Score Weighting (IPW) method \cite{narduzzi2014inverse} or the entropy balancing method \cite{hainmueller2012entropy} to control the impact of confounding factors. Among them, the entropy balancing method has better robustness and more stable weights than the IPW method, so this paper chooses the entropy balancing method as the extension of our research method. However, although the existing entropy balancing method can control the overall impact of confounding factors, it cannot ensure the accuracy of causal effect prediction in each specific time and space. Specifically, in the logistics industry, order distribution is significantly related to time and space factors. If the estimation of causal effects within a specific temporal-spatial domain is inaccurate, it could potentially have adverse (or even negative) impacts on marketing outcomes.

To address the issue that the existing entropy balancing method cannot accurately estimate causal effects in specific temporal-spatial domains, we propose a flexible grid-based temporal-spatial domain partitioning method. In the spatial dimension, a flexible grid based on order volume is used for partitioning, which not only improves the granularity of the partition but also ensures that the order volume within each spatial grid reaches a certain level to avoid sample bias. In the temporal dimension, the partition is based on the distribution of intracity freight order volume. Based on the flexible grid-based temporal-spatial domain partitioning method, we further propose a continuous entropy balancing method (TS-EBCT) in temporal-spatial domain. This method extends the existing continuous entropy balancing method, mainly used to control the impact of confounding variables within the temporal-spatial domain, more accurately estimating the causal effects of the temporal-spatial domain, thereby improving the accuracy of causal effect estimation in the logistics field.

The main contributions of this paper can be summarized into the following three points:
\begin{itemize}
\item This paper proposes a flexible grid-based temporal-spatial domain partitioning method, which not only improves the granularity of the partition but also ensures that the order quantity within each grid reaches a certain threshold.
\item This paper proposes a temporal-spatial domain continuous entropy balancing method (TS-EBCT), which further extends the existing continuous entropy balancing method, solving the problem of how to accurately estimate causal effects under the temporal-spatial imbalance in the logistics field.
\item This paper uses two real datasets constructs two simulation datasets, and conducts detailed experiments and experiments and analyses on these of these four datasets. The experimental results demonstrate the effectiveness and convergence of the proposed TS-EBCTCT method. In terms of eliminating the influence of confounding variables, the TS-EBCT method has obvious advantages over other methods. Moreover, the sample weights obtained by this method are more conducive to the accurate learning of causal effects by the uplift model. We will make the code and datasets publicly available after the paper is published, contributing to the causal inference community.
\end{itemize}
\section{Related Works}
In the field of marketing science, especially in the analysis of causal effects of subsidy and pricing strategies, causal effect learning is mainly conducted on observational data due to the high cost of randomized experiments. However, this method inevitably suffers from confounding bias. To mitigate this impact, \cite{dehejia2002propensity} and \cite{narduzzi2014inverse} proposed the Propensity Score Matching (PSM) and IPW methods to calculate weights, respectively. The main goal is to optimize the covariate balance between the treatment group and the control group, thereby controlling the impact of confounding bias. However, the entropy balancing method proposed by \cite{hainmueller2012entropy} shows three significant advantages over PSM and IPW: 1) Firstly, the entropy balancing method can ensure complete covariate balance between the treatment and control groups, as it directly solves the optimization problem of covariate balance, rather than relying on the random results of PSM or IPW; 2) Secondly, the entropy balancing method does not need to choose a model, as it directly uses the empirical distribution of the observed data, rather than relying on model assumptions; 3) Lastly, entropy balancing method can more effectively use data. Compared with the PSM method, it does not need to discard any data, and compared with the IPW method, it does not over-weight the data, thereby avoiding the efficiency loss of weighted estimation. Entropy balancing method can effectively avoid the generation of extreme weights and has double robustness \cite{zhao2016entropy}. Therefore, entropy balancing method can more effectively control confounding bias and improve the accuracy of causal effect estimation when dealing with the calculation of causal effects of observational data.

In recent years, entropy balancing method has gradually become popular and has been widely applied in practical business scenarios such as e-commerce, freight, and passenger transport, achieving good business benefits. However, in actual use scenarios, binary entropy balancing method cannot meet the requirements. For example, in the subsidy scenario of freight, there are often many types of coupons. Therefore, to solve the continuous treatment scenario, \cite{Tubbicke} proposed a continuous treatment entropy balancing method. This method extends the original entropy balancing method of \cite{hainmueller2012entropy}, introducing the entropy balancing of continuous treatment(EBCT). This continuous entropy balancing method solves a global convex constraint optimization problem and provides an efficient computational implementation method. At the same time, \cite{vegetabile2021} also proposed a non-parametric method for continuous entropy balancing method and provided a user-friendly R package code implementation. In addition, we further extended the continuous entropy balancing method and proposed TS-EBCT method based on a flexible grid, which can eliminate the influence of confounding variables in both time and space dimensions, thereby more accurately learning the causal effects of the temporal-spatial domain. This paper is based on the EBCT method proposed by \cite{Tubbicke,vegetabile2021}, and combined with the stratified regularization entropy balancing method proposed by \cite{xu_yang_2023} for implementation, further avoiding the situation of optimization failure or highly concentrated weights. The specific introduction of this method will be unfolded in the next section.
\section{Methods}
In this section, we will introduce from the following three parts: 1) Firstly, in the first part, we briefly review the optimization process of the previous continuous entropy balancing method, paving the way for the introduction of temporal-spatial continuous entropy balancing later. To illustrate how we partition data in the temporal-spatial domain, the next subsection will provide a more detailed introduction; 2) Secondly, in the second subsection, we will introduce the flexible grid-based temporal-spatial domain partitioning method proposed in this paper. Based on this partitioning method, we further extend the existing continuous entropy balancing method to solve the problem of causal effect estimation under the unbalanced situation in the temporal-spatial domain; 3) Finally, in the third part, we introduce the TS-EBCT method proposed in this paper.
\subsection{Brief Review of EBCT Method}\label{sec:Causal Effects of Continuous Treatments}
Entropy balancing method is one of the most mainstream causal estimation methods at present. In recent years, researchers have further extended the traditional binary entropy balancing method to continuous treatment scenarios to adapt to more practical business scenarios, bringing tremendous business value. Here we briefly outline the work of \cite{fong2018covariate} and \cite{vegetabile2021} in the field of continuous entropy balancing. Suppose $x_i$, $t$, $n$ represent the confounding factor, treatment variable, and sample size, respectively. First, they demean the confounding variables and treatment variables. Without loss of generality, they assume that the means of the confounding variables and treatment variables are 0. Then, the balancing objective is transformed into learning a set of weights $w_{i}, i=1,\dots,n$, such that $\sum_{i=1}^{n} w_{i}x_{i} = 0$,$\sum_{i=1}^{n} w_{i}t_{i} = 0$,$\sum_{i=1}^{n} w_{i}t_{i}x_{i} = 0$. To make the equation expression more concise, here we introduce a 2k+1 dimensional vector $\mathbf{g}= [x_{i}, t_{i}, t_{i}x_{i}]$, where k represents the number of confounding variables after Principal Component Analysis(PCA)\cite{bro2014principal} dimension reduction. Further, the constraint condition can be rewritten as $\sum_{i=1}^{n} w_{i}\mathbf{g} = 0$. The final step is how to solve this optimization problem. \cite{hainmueller2012entropy} proposed minimizing the distribution distance between the weight $w_{i}$ and the base weight $q_{i}$($i=1,\dots,n.$) using KL divergence to minimize the objective function. For more different dispersion measurement methods, you can refer to the work of \cite{wang2020minimal}. Here we choose the commonly used KL divergence function. Using the KL divergence function and balance constraints, the optimization of entropy balance can be formally expressed with the following Eq. (\ref{eq:EBCT function}):
\begin{equation}
\label{eq:EBCT function}
\begin{aligned}
&\widehat{\boldsymbol{w}}=\underset{\boldsymbol{w}}{\operatorname{argmin}} \sum_{i=1}^n w_i \log \left(\frac{w_i}{q_i}\right)\\
&\begin{array}{r@{\quad}r@{}l@{\quad}l}
s.t. &&G \boldsymbol{w}=\mathbf{0},\\
     &&\mathbf{1}^{\top} \boldsymbol{w}=1, \\
     &&w_i \geq 0, \quad \text{for}\quad i=1, \ldots, n.
\end{array}
\end{aligned}
\end{equation}
where $G$ represents the $(2k+1) \times n$ dimensional matrix obtained by stacking $\mathbf{g}$.
The above optimization problem can be solved using the Lagrange dual method,
\begin{equation}
\widehat{\boldsymbol{\lambda}}=\underset{\boldsymbol{\lambda}}{\operatorname{argmin}} \log \left(\mathbf{1}^{\top} \exp \left(-\boldsymbol{\lambda}^{\top} G+\boldsymbol{\ell}\right)\right),
\label{eq:Lagrange dual solution}
\end{equation}
where $\boldsymbol{\ell}_{i}=log(q_{i})$ is the logarithmic value of the base weight. Substituting the solved $\widehat{\boldsymbol{\lambda}}$ into the Eq. (\ref{eq:Lagrange dual solution}), we can get the entropy balance weight $\boldsymbol{w}=softmax(-\boldsymbol{\lambda}^{\top} G+\boldsymbol{\ell})$.
\begin{algorithm}[ht]
\SetAlgoLined
\KwIn{H3 Resolution $\to$ 10 Grids: G, Orders volume in each Grids: Orders.}
\KwOut{Flexible grids with each grid exceeding $2 \%$ of monthly orders.}
\SetKwProg{Fun}{Function}{:}{end}
\Fun{$FQ(G, Orders)$}{
    $FG=$ null\;
    Sort all grids according to order volume: Sorted\_G\;
    \ForEach{$g$ in Sorted\_G}{
        Sorted\_G.remove $(g)$\;
        \If{$\mathrm{g}$ is Not Aggregated}{
            agg\_grid $=g$\;
            \While{Orders in agg\_grid $<2 \%$ of monthly orders}{
                agg\_grid=AggGrid(g,Sorted\_G,orders)\;
            }
            FG.add(agg\_grid)\;
        }
    }
    \Return $F G$\;
}\
\Fun{AggGrid(g, Sorted\_G, orders)}{
    Sort all grids in Sorted\_G according to distance to g: Distance\_g\;
    agg\_grid $=g$\;
    \While{resolution of agg\_grid $<4$}{
        $g^{\prime}=$ Distance\_g.pop()\;
        agg\_grid = aggregate $\left(g, g^{\prime}\right)$\;
    }
    \Return agg\_grid\;
}
\caption{Flexible grid division method.}
\label{alg:Flexible grid division}
\end{algorithm}
\subsection{Flexible Grid-based Temporal-Spatial Domain Partitioning Method}\label{sec:Flexible Grid}
When constructing subsidy and pricing strategies in the field of intracity freight, in order to meet the goal of refinement, the physical partitioning strategy needs to use as small a grid partition as possible. However, overly refined grid partitioning may significantly reduce the order volume within each grid interval, thereby increasing the volatility of the strategy. The geographical distribution of freight orders shows significant non-uniform characteristics. For example, compared to suburban areas, the order density in the location of the logistics market is significantly higher. Therefore, how to automatically and flexibly delineate geographical space based on order density has become a decisive factor affecting the effectiveness of the strategy. Therefore, based on the above phenomena, this paper proposes a new type of flexible grid calculation method. First, the physical space is initialized and subdivided into a grid $\mathcal{G}$ according to the 10-level resolution of H3. Secondly, in order to increase the order density of the grid and reach the monthly order volume threshold, further aggregation will be carried out, and in geography, the flexible grid will still maintain connectivity. In terms of aggregation rules, individual grids will be sorted according to the order volume to obtain the sorted grid $\mathcal{G}_{sorted}$, and then aggregation will start from the grid $\mathcal{G}_{max}$ with the largest order volume. The merging process will prioritize grids that meet the near distance (in a clockwise direction). If the distance is the same, the one with a larger order volume will be prioritized. If the order volume is also the same, a random selection will be made for merging. It is worth noting that the convergence limit of the grid is set to level 4, that is, the side length is 22 kilometers. The specific algorithm flow pseudocode is shown in Algorithm \ref{alg:Flexible grid division}. After partitioning by this method, we can obtain a temporal-spatial domain dataset with flexible grid dimensions. How to more accurately learn the causal effects under the temporal-spatial domain and alleviate the impact of confounding variables, in the next subsection, we will introduce the temporal-spatial Continuous Entropy Balancing method proposed in this paper.

\subsection{Temporal-Spatial Entropy Balancing for Causal Continuous Treatment}\label{sec:Temporal-Spatial Entropy Balancing for Causal Continuous Treatment}
\begin{figure}[ht]
    \centering
    \includegraphics[width=\linewidth]{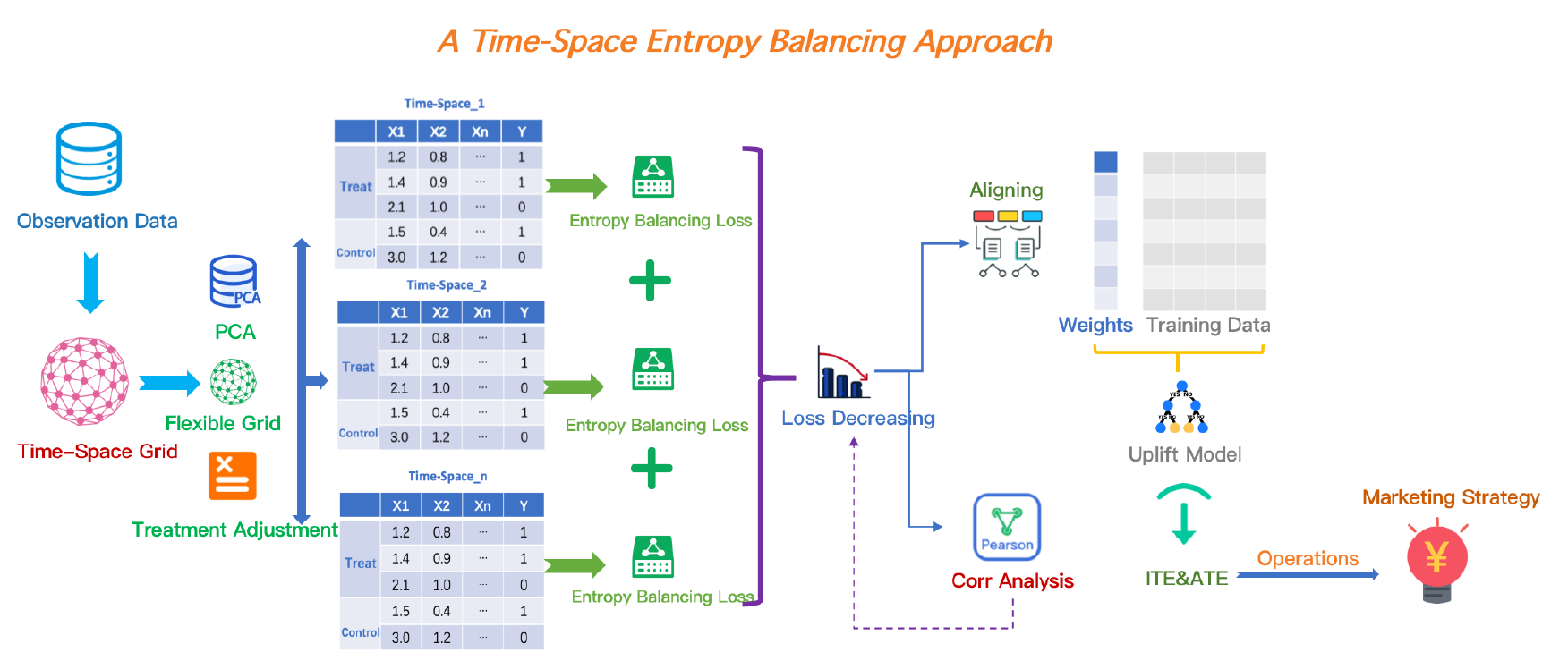}
    \caption{Framework of TS-EBCT method.}
    \label{fig:Framework of TS-EBCT Method}
\end{figure}

In this section, we will provide a detailed introduction to the TS-EBCT method proposed in this paper. The overall architecture of this method in practical business applications is shown in Figure \ref{fig:Framework of TS-EBCT Method}. Firstly, we divide the observed data into different temporal-spatial domain sub-intervalservals according to the temporal-spatial or flexible grid. The goal of this division method is to ensure that the number of samples in each grid reaches a preset threshold, while refining the grid division as much as possible to better reflect the characteristics of temporal-spatial data. Next, to deal with the explosion of feature dimensions, we use PCA to reduce the dimensionality of the features. The dimension of reduction is a hyperparameter, which can be determined by the explained variance score, choosing to retain sufficient information. After feature dimension reduction, we can calculate the Entropy Balancing Loss in each temporal-spatial grid, which is used to evaluate the imbalance degree of confounding variables in each grid. Entropy balancing loss is calculated using KL divergence, which can measure the correlation between confounding variables and treatment variables. Further, by summarizing the entropy balancing loss of each temporal-spatial grid, we can consider the confounding problem in the entire temporal-spatial domain. Finally, we can use optimization methods such as gradient descent to update the weight parameters to be solved, in order to minimize the total loss function. This optimization process aims to find a set of weight values that can balance the impact of each impact of each sample in the overall and temporal-spatial dimensions, thereby reducing the interference of confounding variables on the estimation of causal effects.
Due to the influence of confounding bias in observation data, the accuracy of causal effect calculation will be affected. To control this influence, we propose the TS-EBCT method, which controls the influence of confounding variables in the control group and treatment group by learning a new set of sample weights $\boldsymbol{w}$, thereby improving the accuracy of causal effect estimation. The temporal-spatial domain entropy balancing weight $\boldsymbol{w}$ is obtained in the following way:
\begin{equation}
\label{eq:spatio-temporal entropy balance weight}
\begin{aligned}
&\min_{w} H(w)=\sum_{i}\sum_{j} h\left(w_{ij}\right)\\
&\begin{array}{r@{\quad}r@{}l@{\quad}l}
s.t. &&\sum_{i} \sum_{j}\left\|\overline{X}_{ij}-{X_{ij}}^{T}_{c} w_{ij}\right\|_2^2=0\\
     &&\sum_i\sum_j w_{ij}=1, w \ge 0, \\
\end{array}
\end{aligned}
\end{equation}
Here, $h(\cdot)$ represents any distance metric function, specifically, this article uses the KL divergence function. In the actual calculation process, for convenience of calculation, the $\sum_{i} \sum_{j}\left\|\overline{X}_{ij}-{X_{ij}}^{T}_{c} w_{ij}\right\|_2^2=0$ in Eq. (\ref{eq:spatio-temporal entropy balance weight}) is transformed. It is transformed into the mean of the feature $f$ of the control group within the spatio-temporal grid $ij$ being equal to the mean of the treatment group. The transformed constraint condition is shown in Eq. (\ref{eq:constraint constraint condition}).
\begin{equation}
\label{eq:constraint constraint condition}
\begin{aligned}
\sum_{i}\sum_{j} w_{ij}{c_{f}}_{ij}\left(X_{ij}\right)={t_{f}}_{ij}
\end{aligned}
\end{equation}
where $f\in 1,\dots,F$, $F$ represents the number of features; $i\in 1,\dots,I$, $I$ represents the number of time dimensions; $j\in 1,\dots,J$, $J$ represents the number of spatial dimensions; ${c_{f}}_{ij}$ is the sum of the columns of the control group with time dimension $i$, spatial dimension $j$, and feature $f$; ${t_{f}}_{ij}$ is the sum of the columns of the treatment group with time dimension $i$, spatial dimension $j$, and feature $f$. The above problem can transform into a Lagrange dual problem, and the transformed expression is shown in Eq. (\ref{eq:Lagrange optimization}),
\begin{equation}
\label{eq:Lagrange optimization}
\begin{aligned}
\min_{w, \lambda_0, Z} L^p &=\sum_{i}\sum_{j} w_{ij} \log \left(\frac{w_{ij}}{q_{ij}}\right)\\
&+\sum_{i}\sum_{j}\sum_f \lambda_{fij}\left(\sum_{i}\sum_{j} w_{ij} {c_{f}}_{ij}\left(X_{ij}\right)-{t_{f}}_{ij}\right)\\
&+\left(\lambda_0-1\right)\left(\sum_{i}\sum_{j} w_{ij}-1\right),
\end{aligned}
\end{equation}
\begin{algorithm}[ht]
\SetAlgoLined
\KwIn{X, T, OD. \tcp{X is features, T is Treatment, OD is the flexible grid.}} 
\KwOut{$\boldsymbol{w}$. \tcp{sample weights.}} 
\SetKwProg{Fun}{Function}{:}{end}
\Fun{TSEBCT(X, T, OD)}{
    X = PCA(X); \tcp{using PCA method to decrease feature dimensions.}
    T = PolynomialFeatures(T, degree=1); \tcp{generate higher moments.}
    T = Standardization(T); \tcp{standardization data.}
    X\_mean = null\;
    \ForEach{$od$ in $OD$}{  
        X\_temp = X[X==od]\; 
        X\_temp = Standardization(X\_temp)\;
        \If{X\_mean == null}{
            X\_mean = X\_temp\;
        }
        \ElseIf{X\_mean != null}{
            X\_mean = Column\_stack(X\_mean, X\_temp)\;
        }
    }
    gTx\_int = Multiple(X\_mean, T[:,0])\;
    gTx = Column\_stack(T, X\_mean, gTx\_int)\;
    tr\_total = gTx.sum(); \tcp{constrain matrix}
    tr\_total[0] = 1;  \tcp{constrains the sum of weights to 1.}
    tr\_total[-gTx\_int.shape[1]:] = 0;  \tcp{constrains cross item value to 0.}
    $\boldsymbol{w}$ = Optimization(tr\_total, gTx, base\_weight); \tcp{using newton's method.}
    \Return $\boldsymbol{w}$\;
}
\caption{Pseudocode of TS-EBCT method.}
\label{alg:Pseudocodeocode of TS-EBCT Method}
\end{algorithm}
where Z is the Lagrange operator, $Z=\left\{{\lambda_{1}}_{1,1}, \ldots,{\lambda_{F}}_{I,J}\right\}^{\prime}$, since $L^p$ is a strictly convex function, assuming its derivative $\cfrac{\partial L^p}{\partial w_{ij}}=0$, the solution can be obtained as $w_{ij}^*=\frac{q_{ij} \exp \left(-\sum_{f=1}^F \lambda_{fij} c_{fij }\left(X_{ij}\right)\right)}{\sum_i\sum_j  q_{ij} \exp \left(-\sum_{f=1}^F \lambda_{fij} c_{fij}\left(X_{ij}\right)\right)}$. Next, bring the solution back to the Eq. (\ref{eq:Lagrange optimization}) can obtain the following Eq. (\ref{eq:Optimization function}):
\begin{equation}
\label{eq:Optimization function}
\begin{aligned}
\min _Z L^d=&\log \left(\sum_i\sum_j q_{ij} \exp \left(-\sum_{f=1}^F {\lambda_{f}}_{ij} {c_{f}}_{ij}\left(X_{ij}\right)\right)\right)\\
&+\sum_{i}\sum_{j}\sum_{f} {\lambda_{f}}_{ij} {t_{f}}_{ij},
\end{aligned}
\end{equation}
where $Z=\left\{{\lambda_{1}}_{1,1}, \ldots, {\lambda_{F}}_{I,J}\right\}^{\prime}$ displayed in vector form can get the following Eq. (\ref{eq:loss function}):
\begin{equation}
\label{eq:loss function}
\min _Z L^d=\log \left(Q^{\prime} \exp \left(-C^{\prime} Z\right)\right)+M^{\prime} Z,
\end{equation}
where matrix $C = [c_{1}(X_{ij}), \dots, c_{F}(X_{ij})]'$, $M = [{t_{1}}_{ij}, \dots, {t_{F}}_{ij}]'$.
The solution to Eq. (\ref{eq:loss function}) is $w^*=\frac{Q \cdot \exp \left(-C^{\prime} Z\right)}{Q^{\prime} \exp \left(-C^{\prime} Z\right)}$, $\quad Q=\left[q_{0}, \ldots, q_{n}\right]^{\prime}$.
Finally, we can solve Z according to the gradient descent method (e.g., Newton's method), the iterative calculation process is as follows:
\begin{equation}
Z^{\text {new }}=Z^{\text {old }}-l \nabla_Z^2 L^{d^{-1}} \nabla_Z L^d,
\end{equation}
where gradient=$\frac{\partial^2 L^d}{\partial Z^2}=C\left[D(\boldsymbol{w})-\boldsymbol{w} {\boldsymbol{w}}^{\prime}\right] C^{\prime}$, learning rate is $l$, $D(\boldsymbol{w})$ is the diagonal matrix of $\boldsymbol{w}$.
As shown in Algorithm \ref{alg:Pseudocodeocode of TS-EBCT Method}, it is the execution process of the entire algorithm. After obtaining the sample weights through the TS-EBCT method, we will evaluate the algorithm: 1) Firstly, the obtained weights are used to weight the samples; 2) Then, calculate the correlation between each feature and the treatment variable; 3) Finally, compare the changes in the two correlations under the weighted and unweighted conditions, and whether the correlation has decreased. After the algorithm evaluation, we will use this sample weight to learn the causal effect in the Uplift model. Specifically, taking the single model(S-learner) model as an example, we use the entropy balancing weight as the sample weight of the S-learner base model (e.g., lightgbm). After learning the causal effect, we will use integer programming methods to solve specific subsidy or pricing strategies.
\section{Experiments}
In this section, we will introduce the experimental datasets, evaluation methods, and experimental results separately. Through experimental comparisons on different datasets, we demonstrate the effectiveness and robustness of the temporal-spatial domain entropy balance proposed in this paper. Experimental results show that our method outperforms the baseline approach in eliminating the correlation between covariates and treatment variables. In addition, the sample weights learned by our method can be applied to causal inference models to significantly improve the area under the uplift curve(AUUC) \cite{gutierrez2017causal} metric.
\subsection{Data Sets}
In this paper, we construct two synthetic data sets and use two real data sets from subsidy and pricing business. The construction method of the synthetic data sets and the source and distribution of the real data sets will be introduced in detail below.
\subsubsection{Synthetic DataSets}
To demonstrate the effectiveness of our proposed method, we constructed synthetic datasets for validation. We set the dataset size to $n$ = 100000 and the dimension of the observed variables to $p$ = 100. The observed variables $X=[x_1, x_2, \dots, x_p]$ were generated from independent Gaussian distributions as follows:
\begin{equation}
x_1,x_2,\dots,x_p \sim N(0,1),
\end{equation}
where $x_i$ represents the $i$-th value of the observed variable $X$.
To simulate the covariate imbalance problem that exists in data from different temporal-spatial dimensions, we have added a new feature that represents the spatial domain in which the data is located. This feature is generated by the following binomial distribution:
\begin{equation}\label{OD}
    OD \sim Binomial(100,0.9).
\end{equation}
We made some modifications to the data construction method of \cite{kuang2017estimating} to generate multiple treatment variable $T$. First, we standardized the generated variable to map it to interval [0,1] and then to generate enough control group data, we shifted the values of the treatment variable. The specific generation method is as follows:
\begin{equation}
   T_{missp} = \left\{\begin{split}&t-0.4, \quad t>0.4\\
   &0, \quad t\leq 0.4.
\end{split}\right.
\end{equation}
where $t$ is defined as $t=Norm({\textstyle \sum_{i=1}^{p\cdot r_c}} \ s_c\cdot x_i+N(0,1))$, the confounding rate $r_c$ and the confounding strength $s_c$ both fall within the interval [0,1]. The confounding rate represents the ratio of confounding factors to all observed variables, while the confounding strength refers to the biasing intensity of confounding factors on the treatment.

\begin{figure}[ht]
\centering
\includegraphics[width=0.9\linewidth]{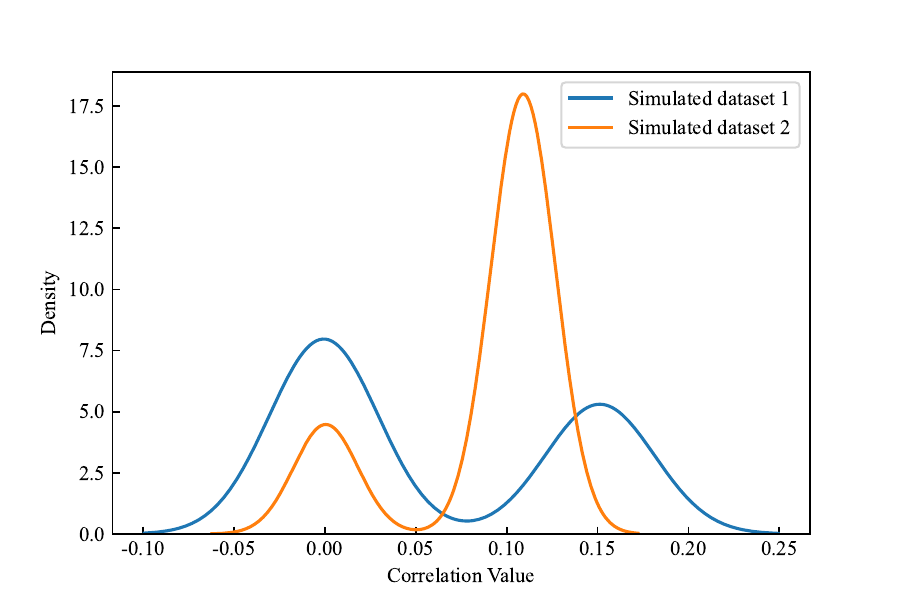}
\caption{Correlation distribution between features and treatment in simulated datasets.}
\label{fig:Correlation Distribution between Features and treatment in Simulated Datasets}
\end{figure}

We generate the feedback variable $Y$ using a linear function, and the generation method is as follows:
\begin{equation}
\begin{aligned}
Y_{linear} &= T + {\textstyle \sum_{j=1}^{p}\left \{ I(mod(j,2)\equiv0 )\cdot(\frac{j}{2} +T)\cdot x_j  \right \} }\\
&+N(0,3),
\end{aligned}
\end{equation}
where $I(\cdot)$ represents the indicator function, and the function $mod(x, y)$ denotes the modulus of $x$ with respect to $y$. We set $s_c$ at 0.5 and generated two simulated datasets with different $r_c$ set at 0.4 and 0.8, respectively. The specific data descriptions are presented in Table \ref{table:Description of simulation datasets}. Additionally, in Figure \ref{fig:Correlation Distribution between Features and treatment in Simulated Datasets}, we show the distribution of the correlation between features and the treatment variable for both simulated datasets. For the dataset with a higher confounding rate, simulated dataset2, a higher proportion of features are correlated with the treatment variable.
\begin{table}[ht]
\caption{Summary of synthetic data sets.}
\label{table:Description of simulation datasets}
\centering
\resizebox{\linewidth}{!}{
\begin{tabular}{|c|c|c|c|c|c|}
\hline
dataset name & sample size & OD size & $s_{c}$ & $r_{c}$ & positive-negative ratio \\ \hline
simulated dataset 1 & 10w & 25 & 0.5 & 0.4 & 1:1 \\ \hline
simulated dataset 2 & 10w & 26 & 0.5 & 0.8 & 1:1 \\ \hline
\end{tabular}
}
\end{table}
\subsubsection{Real DataSets}
To further validate the effectiveness of our proposed method in real-world business scenarios, we carried out experimental tests and validation using a dataset of freight user subsidies and pricing.
\paragraph{\textbf{Subsidy dataset.}} This is real dataset 1, which comes from the data of the freight user subsidy scenario. It comprises about 200000 samples and 100 features, and the number of OD is 427. The treatment variable is a continuous variable, and the ratio of positive to negative samples is 1:3.
\paragraph{\textbf{Pricing dataset.}} This is real dataset 2, which comes from the freight user surcharge scenario. It contains 30 features, including about 160000 samples and the OD's count is 377. The treatment variable is a continuous variable and the positive and negative sample ratio is 1:1.5.
\begin{figure}[ht]
\centering
\includegraphics[width=0.9\linewidth]{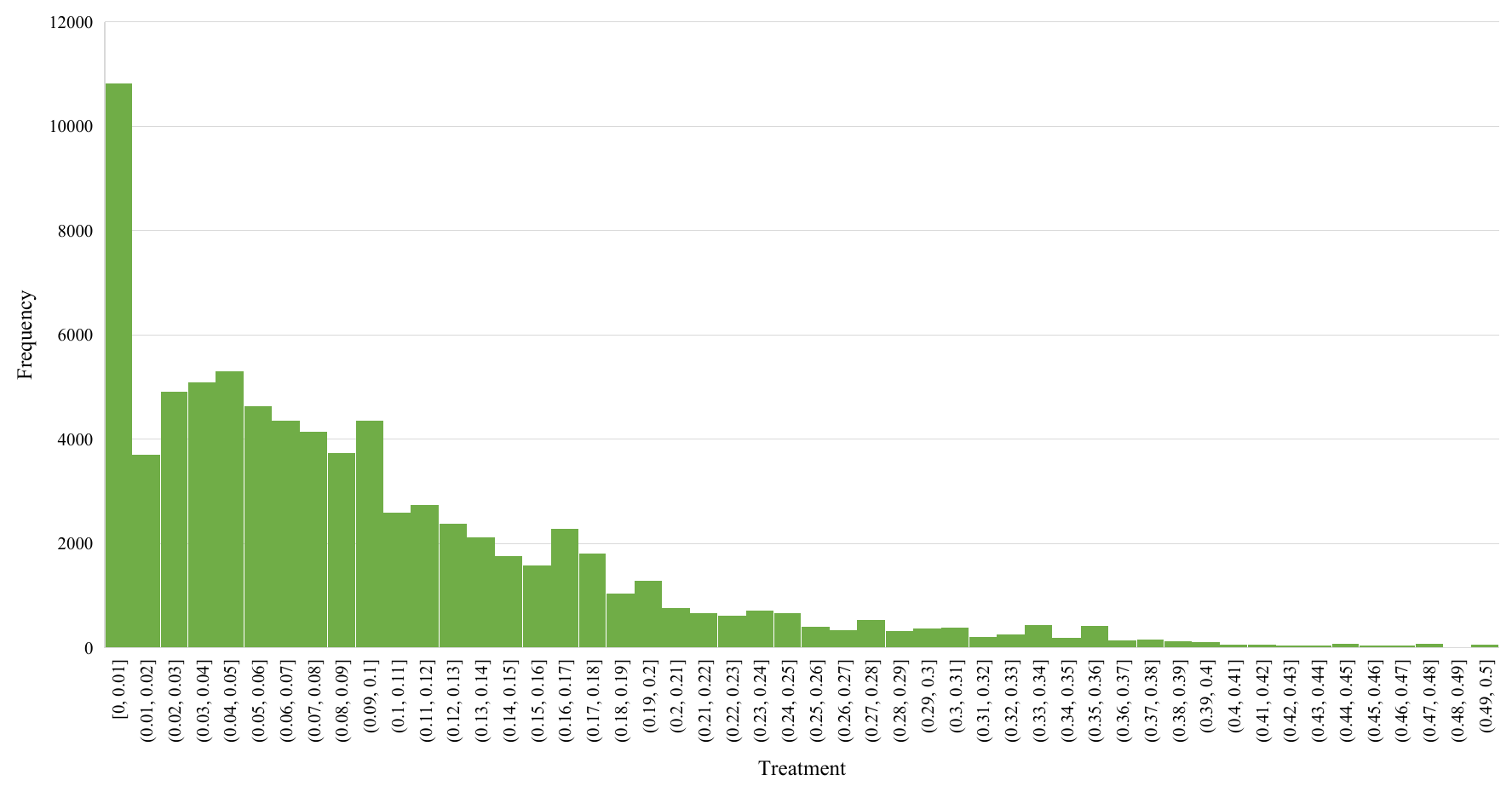}
\caption{Treatment distribution of real dataset 1.}
\label{fig:Treatment distribution of Real Dataset1}
\includegraphics[width=0.9\linewidth]{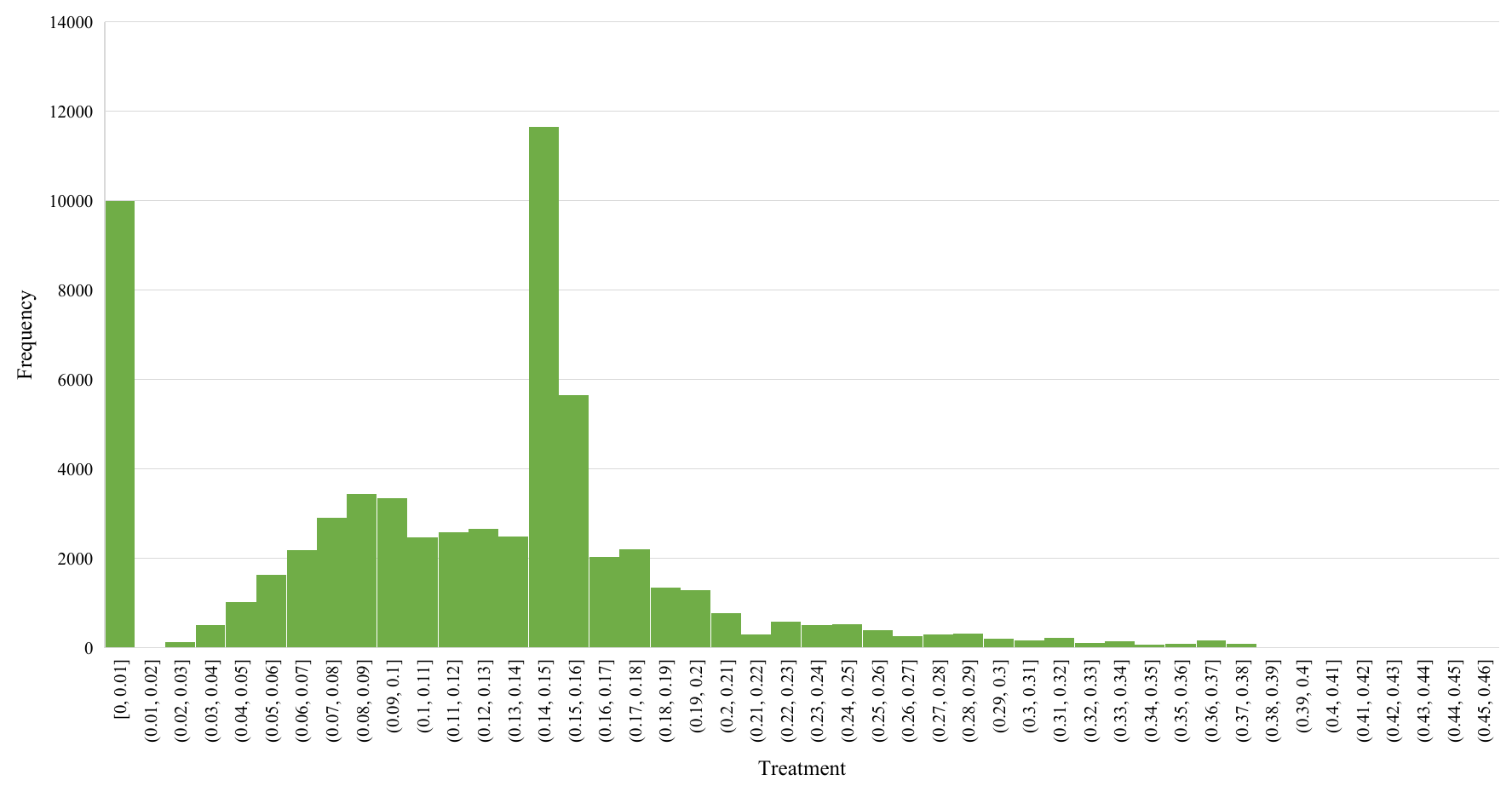}
\caption{Treatment distribution of real dataset 2.}
\label{fig:Treatment distribution of Real Dataset2}
\end{figure}
Table \ref{table:Description of real datasets} provides a brief summary of the two real datasets. Figure \ref{fig:Treatment distribution of Real Dataset1} and Figure \ref{fig:Treatment distribution of Real Dataset2} represent the distribution of treatments in the two real datasets. In real dataset 1, the treatment distribution exhibits a long-tail shape, while in real dataset 2, the treatment distribution resembles a normal distribution.
\begin{table}[ht]
\caption{Summary of real datasets.}
\label{table:Description of real datasets}
\centering
\resizebox{\linewidth}{!}{
\begin{tabular}{|c|c|c|c|c|c|}
\hline
dataset name & number of features & sample size & OD size & treatment type & positive-negative ratio \\ \hline
real dataset 1 & 100 & 20w & 427 & continuous & 1:3 \\ \hline
real dataset 2 & 30 & 16w & 377 & continuous & 1:1.5 \\ \hline
\end{tabular}
}
\end{table}
\subsection{Baseline Methods}
In this section, we will briefly introduce the experimental comparison evaluation method. To fairly evaluate our method and verify the effectiveness of the proposed approach, we will compare TS-EBCT with other re-weighting methods.
\begin{itemize}
\item\textbf{IPW}. \cite{narduzzi2014inverse} proposed using ordinary least squares(OLS) method to solve for weights. IPW method often leads to unstable balance and sometimes even cause an increase in imbalance.
\item\textbf{EBCT}. \cite{Tubbicke} proposed a continuous treatment entropy balance method, EBCT. EBCT removes the correlation between the covariates and continuous treatment variables and effectively avoids assigning extreme weights to individual samples. Additionally, compared to other weighting methods, EBCT has a lower bias and variance.
\item\textbf{TS-EBCT (Ours)}. The TS-EBCT method proposed in this paper further extends the EBCT method, enabling it to maintain entropy balance in both the temporal-spatial domain and overall.
\end{itemize}
\subsection{Results}
This section mainly verifies the effectiveness and convergence of the TS-EBCT algorithm proposed in this paper by comparing experimental results on two synthetic datasets and two real datasets. It is mainly divided into the following three parts: 1) The first part is to verify whether the algorithm proposed in this paper converges by plotting loss curves; 2) The second part mainly compares the weights obtained by different re-weighting methods to eliminate the correlation between covariates and treatment variables; 3) The third part mainly verifies the causal effect promotion effect and its size obtained by different re-weighting methods.
\begin{figure}[ht]
    \centering
    \subfigure[Simulated dataset 1]{
        \includegraphics[width=0.45\linewidth]{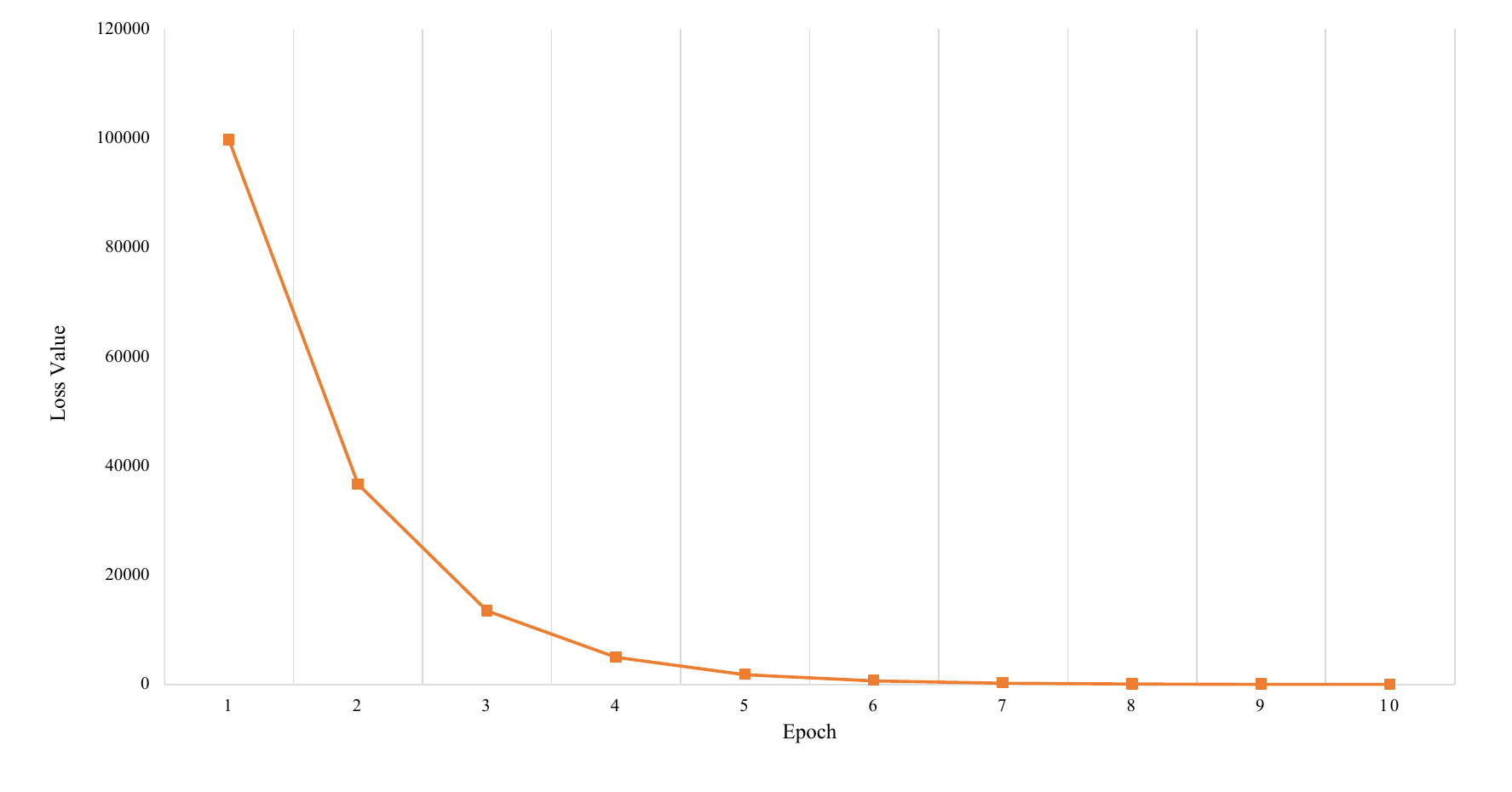}
    }
    \subfigure[Simulated dataset 2]{
	\includegraphics[width=0.45\linewidth]{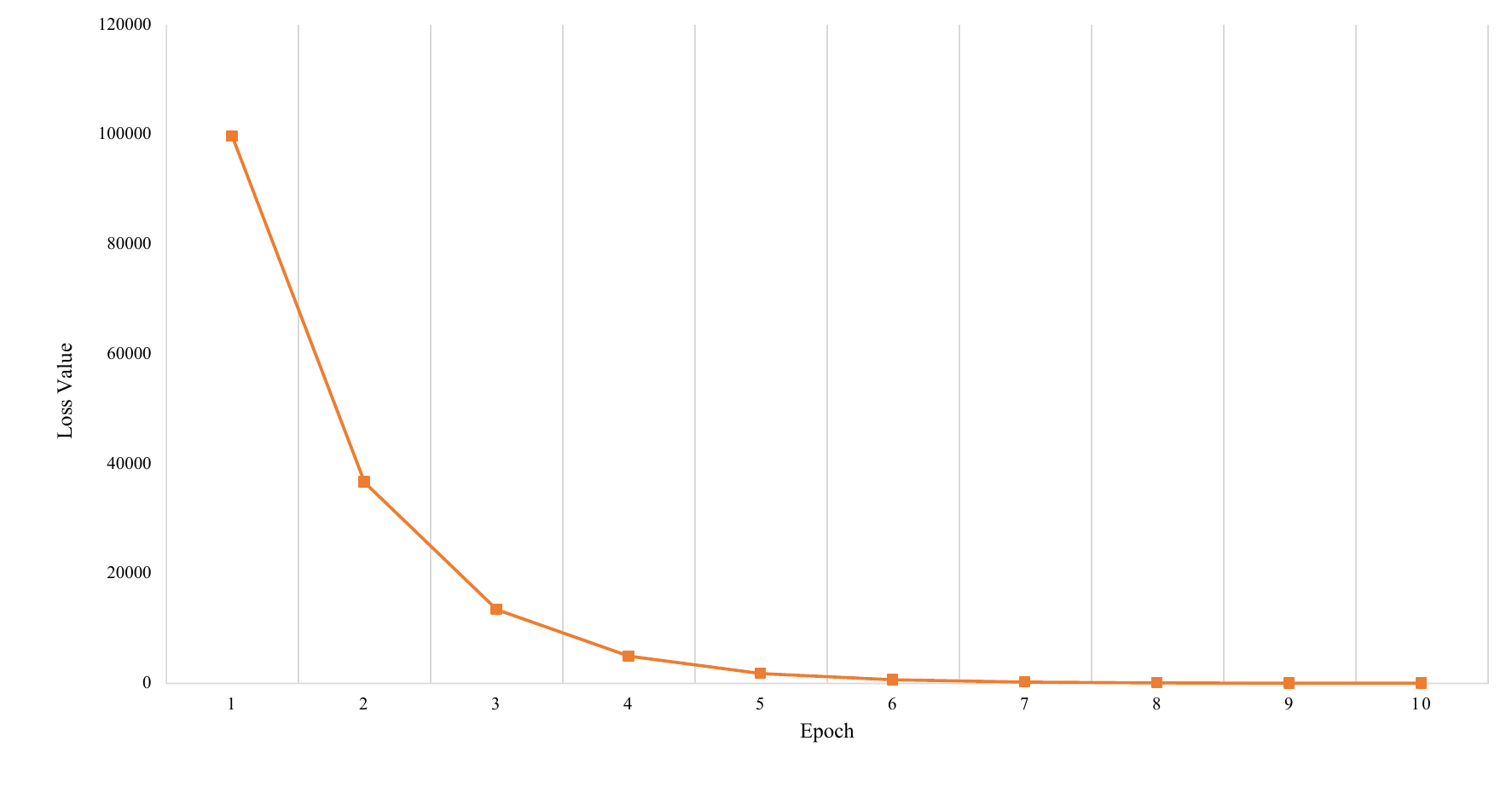}
    }
    \quad
    \subfigure[Real dataset 1]{
        \includegraphics[width=0.45\linewidth]{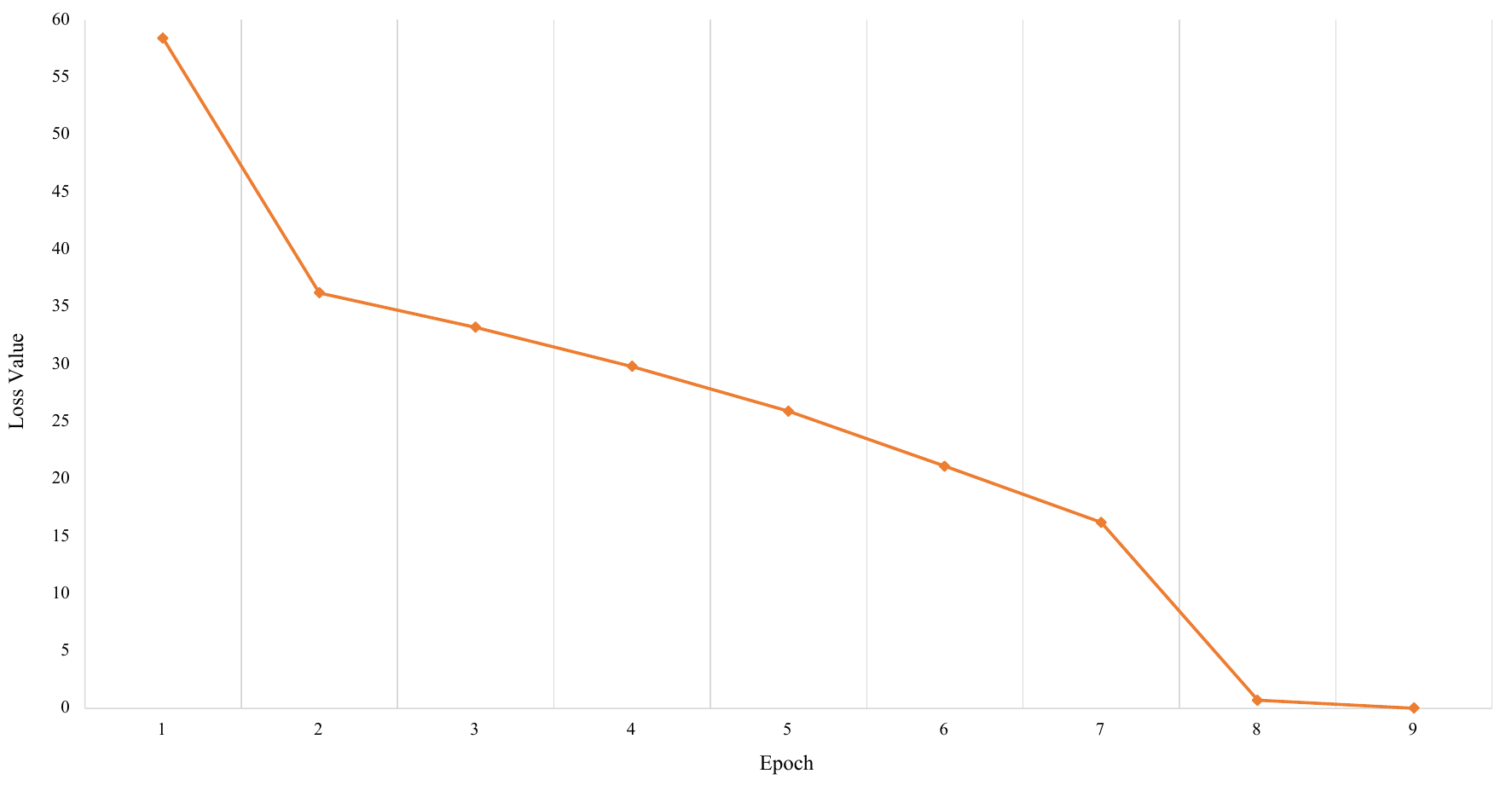}
    }
    \subfigure[Real dataset 2]{
	\includegraphics[width=0.45\linewidth]{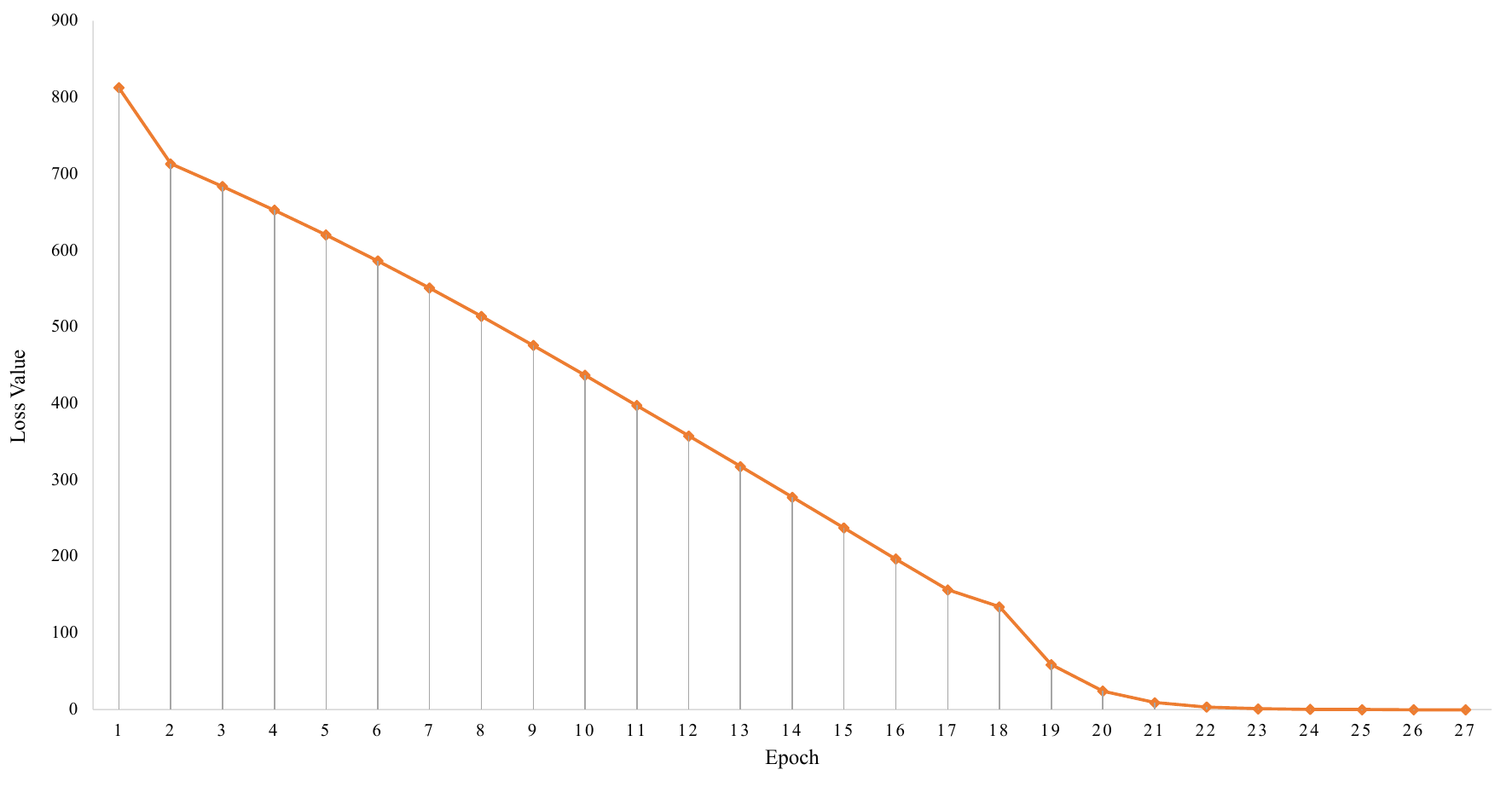}
    }
    \caption{Loss curves on various datasets.}
    \label{fig:loss curve}
\end{figure}
\subsubsection{Loss Curves}
To verify the convergence of TS-EBCT method, experiments were conducted on two simulated and two real datasets. As shown in Figure \ref{fig:loss curve}, the convergence of the TS-EBCT method proposed in this paper is shown on the four datasets. The loss value approaching 0 indicates better convergence, and the faster the loss value decreases, the faster the convergence speed. On the simulated datasets, the loss curves converge more smoothly, with faster convergence at the beginning and gradually stabilizing later, as the loss approaches 0. On the two real datasets, the loss has converged at a relatively small number of epochs, with about 10 epochs on real dataset 1 and about 23 epochs on real dataset 2. During the model training process in this paper, we set the learning rate to 1, and the convergence condition for the algorithm is whether the loss is less than 0.01. The specific experimental parameters can be adjusted according to the actual situation.
\subsubsection{Performance Balancing Covariates}
\begin{figure}[ht]
    \centering
    \includegraphics[width=0.45\linewidth]{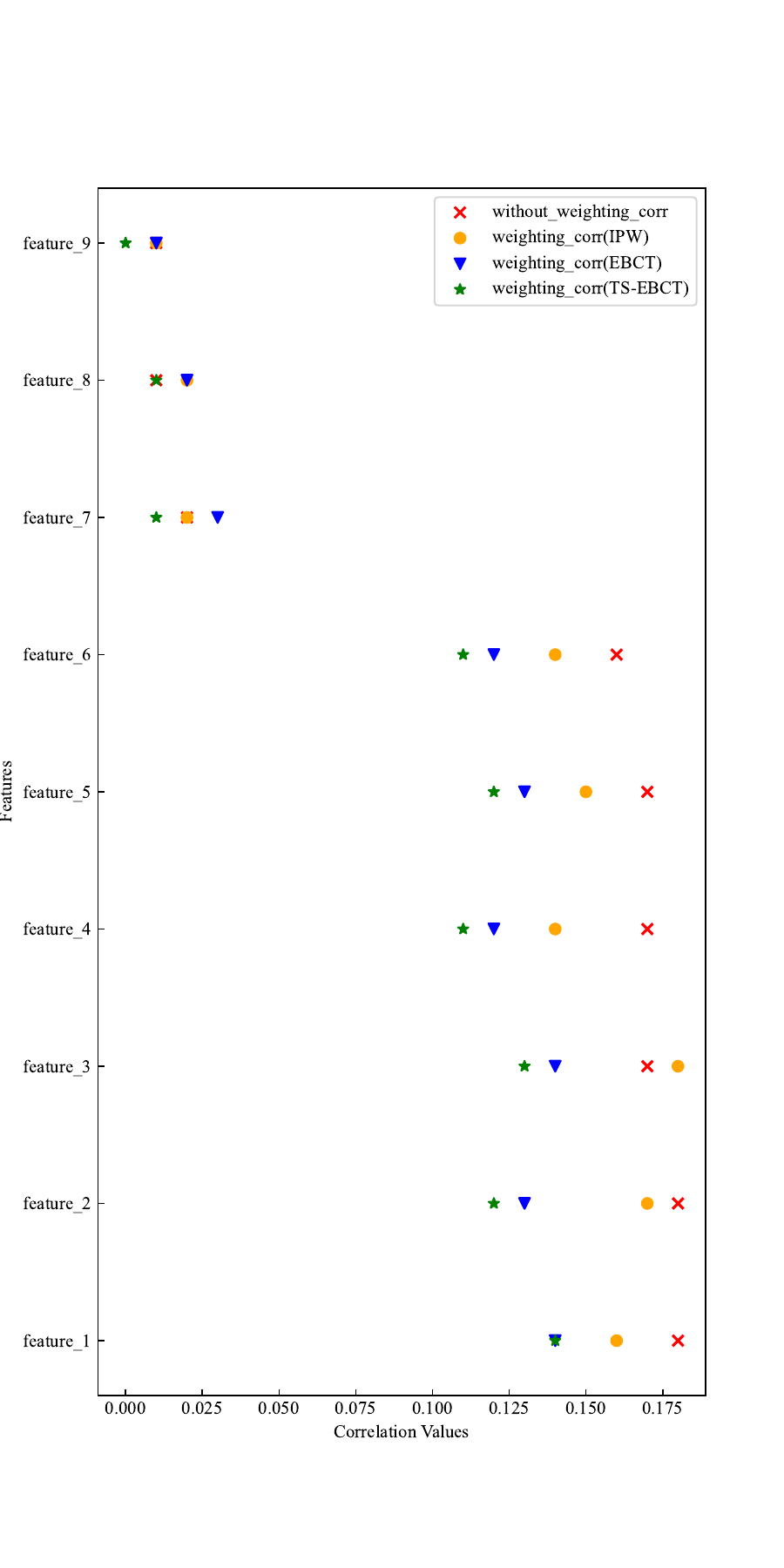}
    \includegraphics[width=0.45\linewidth]{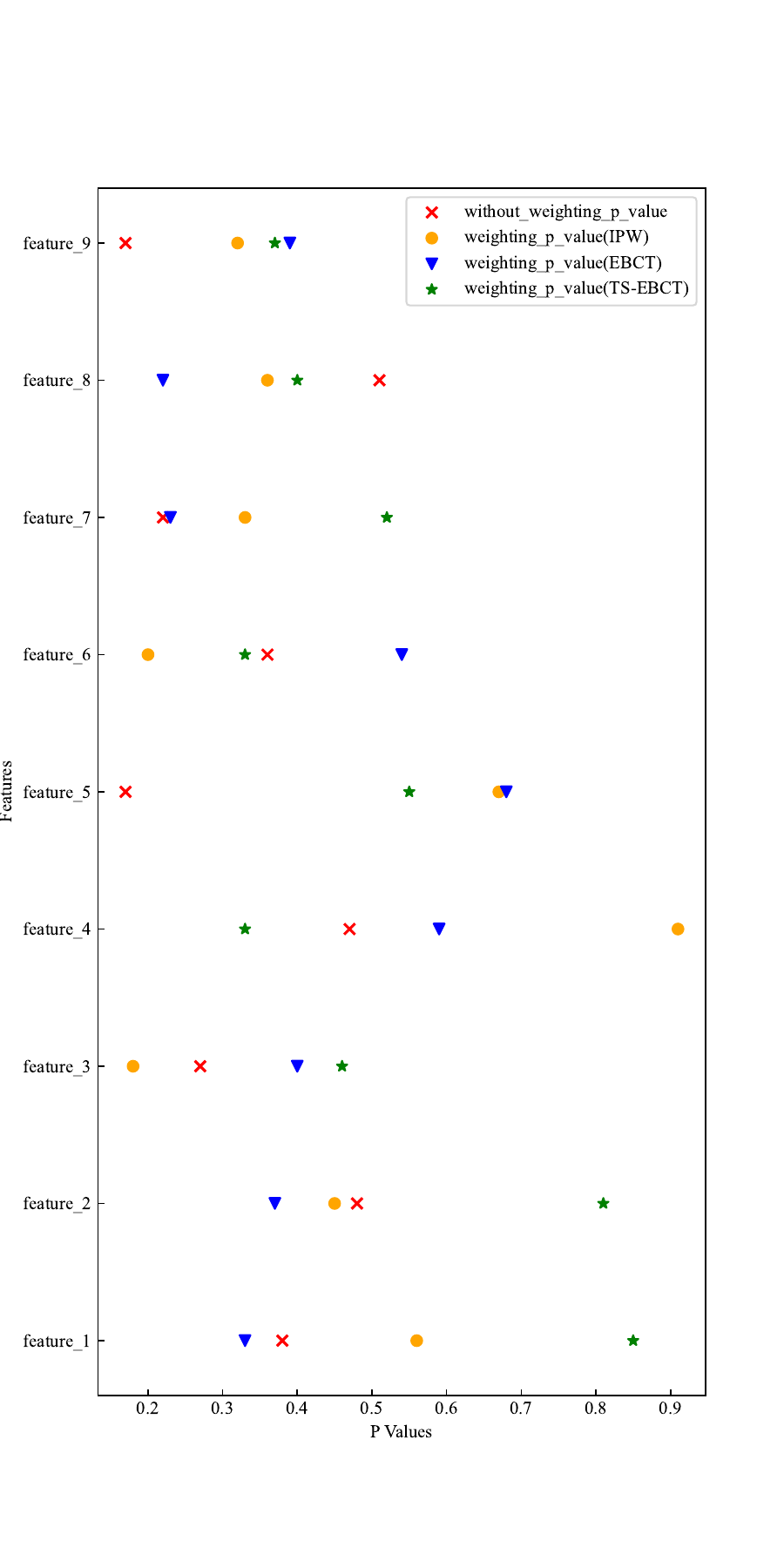}
    \caption{The feature elimination results of simulated dataset 1.}
    \label{fig:The feature elimination results of simulation dataset1}
\end{figure}

\begin{figure}[ht]
    \centering
    \includegraphics[width=0.45\linewidth]{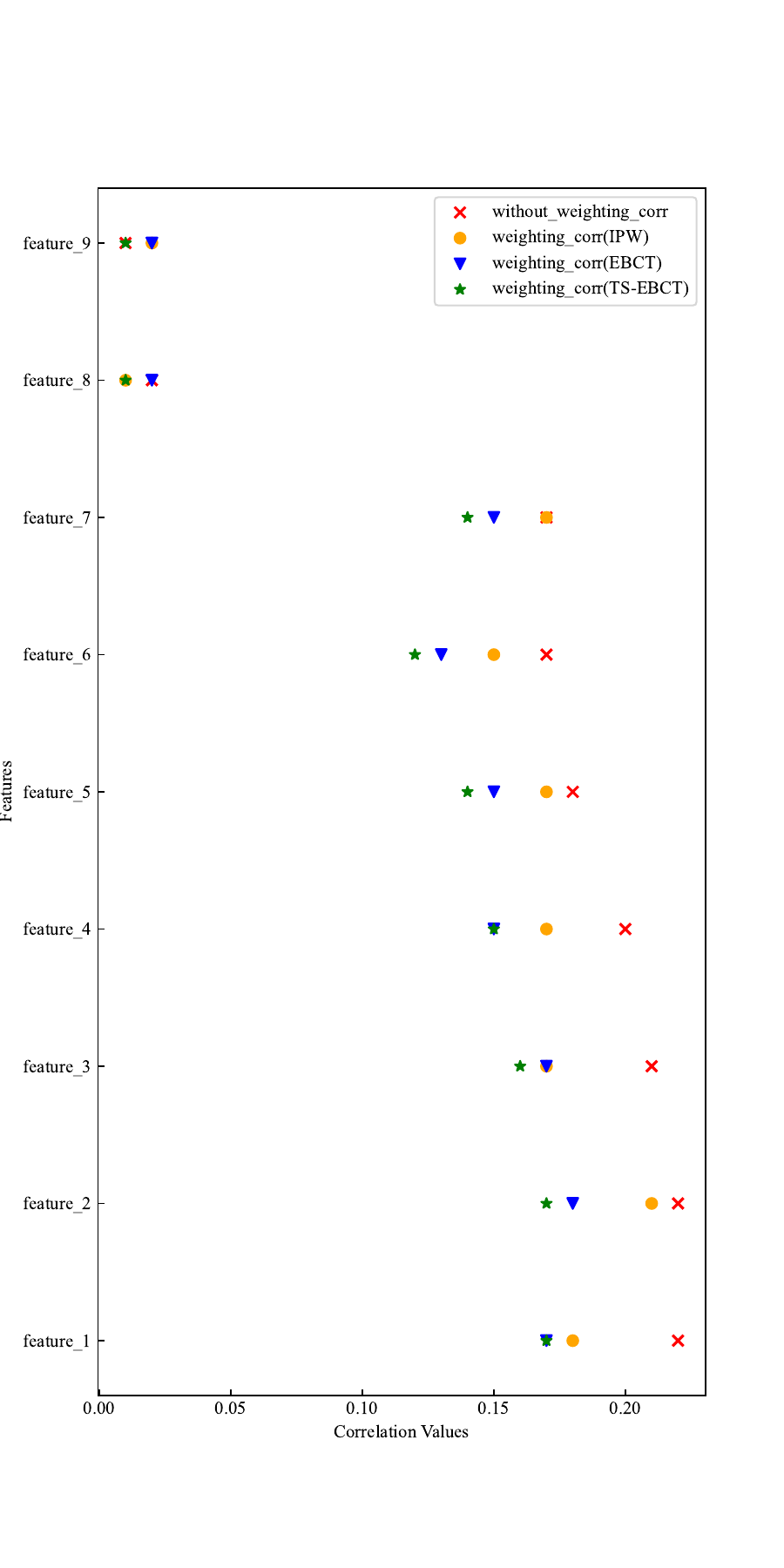}
    \includegraphics[width=0.45\linewidth]{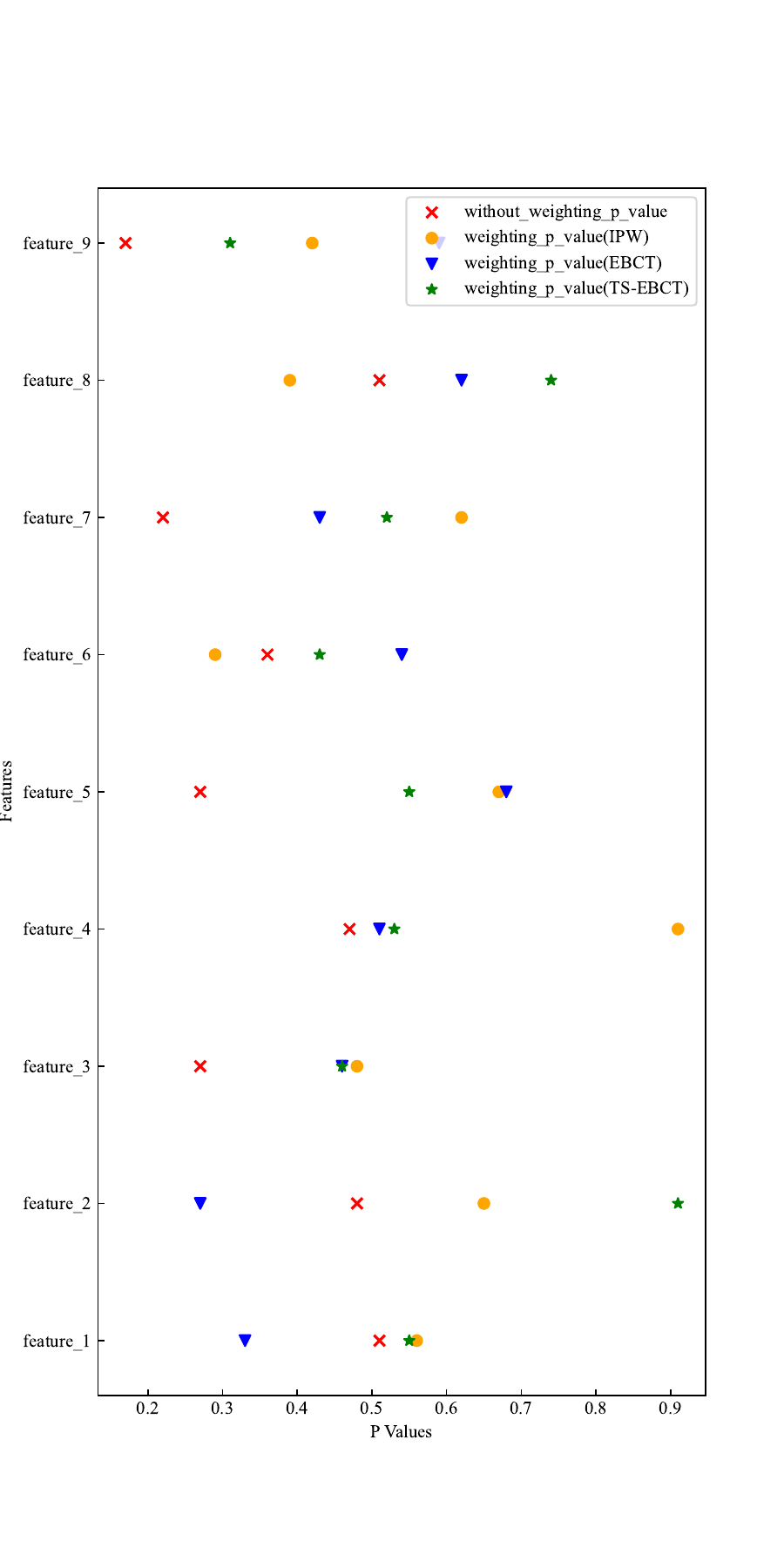}
    \caption{The feature elimination results of simulated dataset 2.}
    \label{fig:The feature elimination results of simulation dataset2}
\end{figure}

\begin{figure}[ht]
    \centering
    \includegraphics[width=0.45\linewidth]{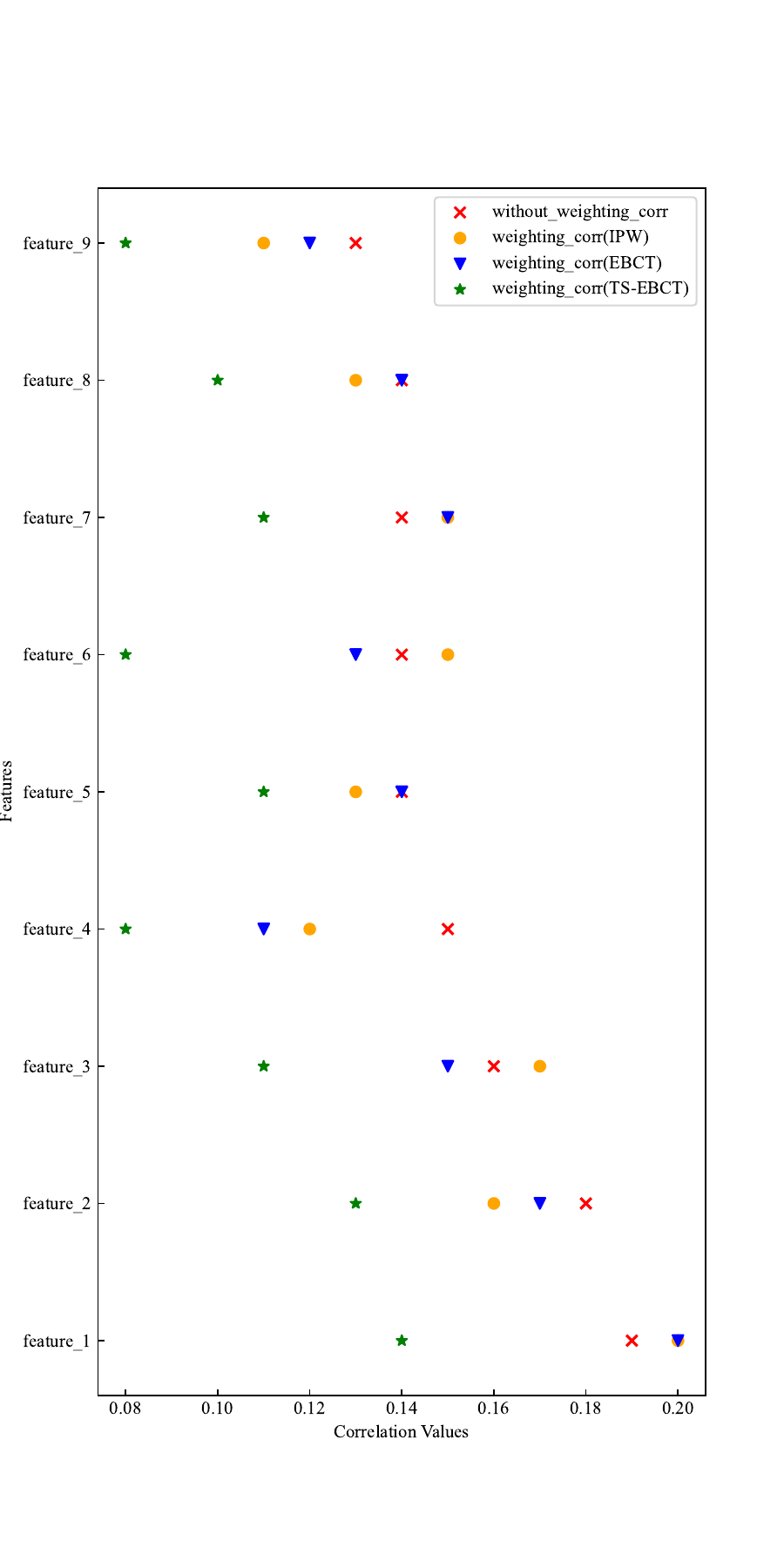}
    \includegraphics[width=0.45\linewidth]{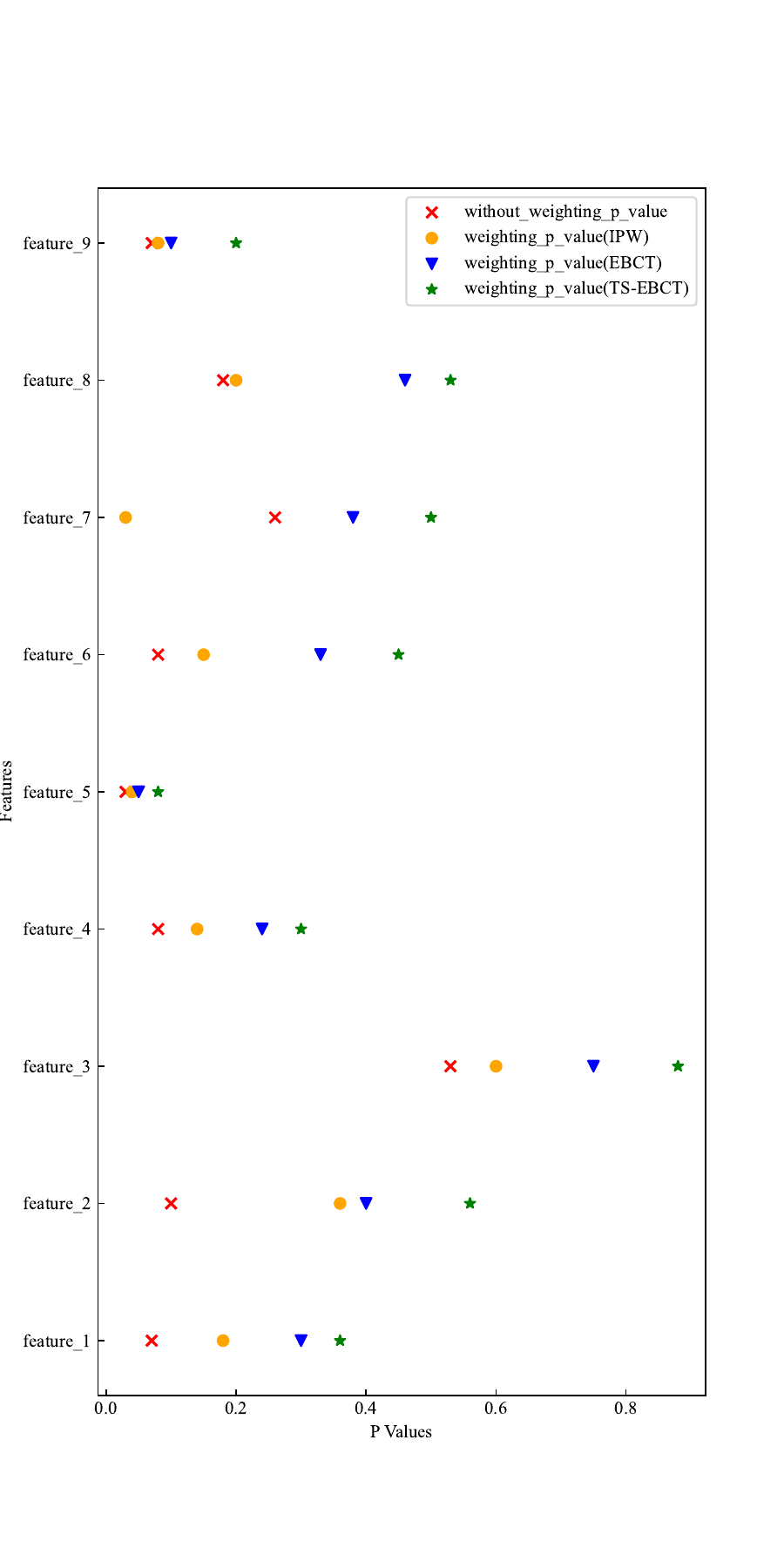}
    \caption{The feature elimination results of real dataset 1.}
    \label{fig:The feature elimination results of real dataset1}
\end{figure}

\begin{figure}[ht]
    \centering
    \includegraphics[width=0.45\linewidth]{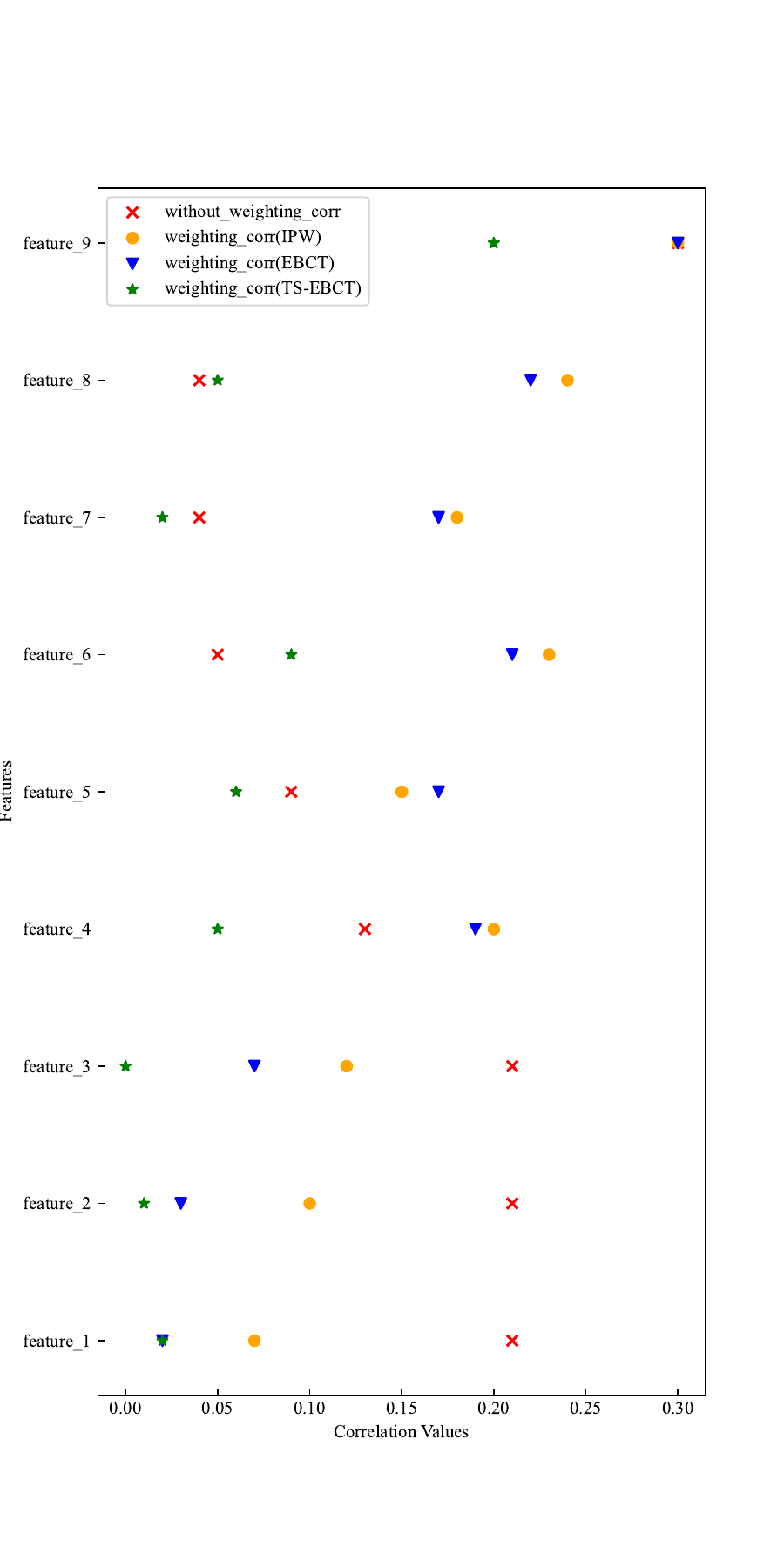}
    \includegraphics[width=0.45\linewidth]{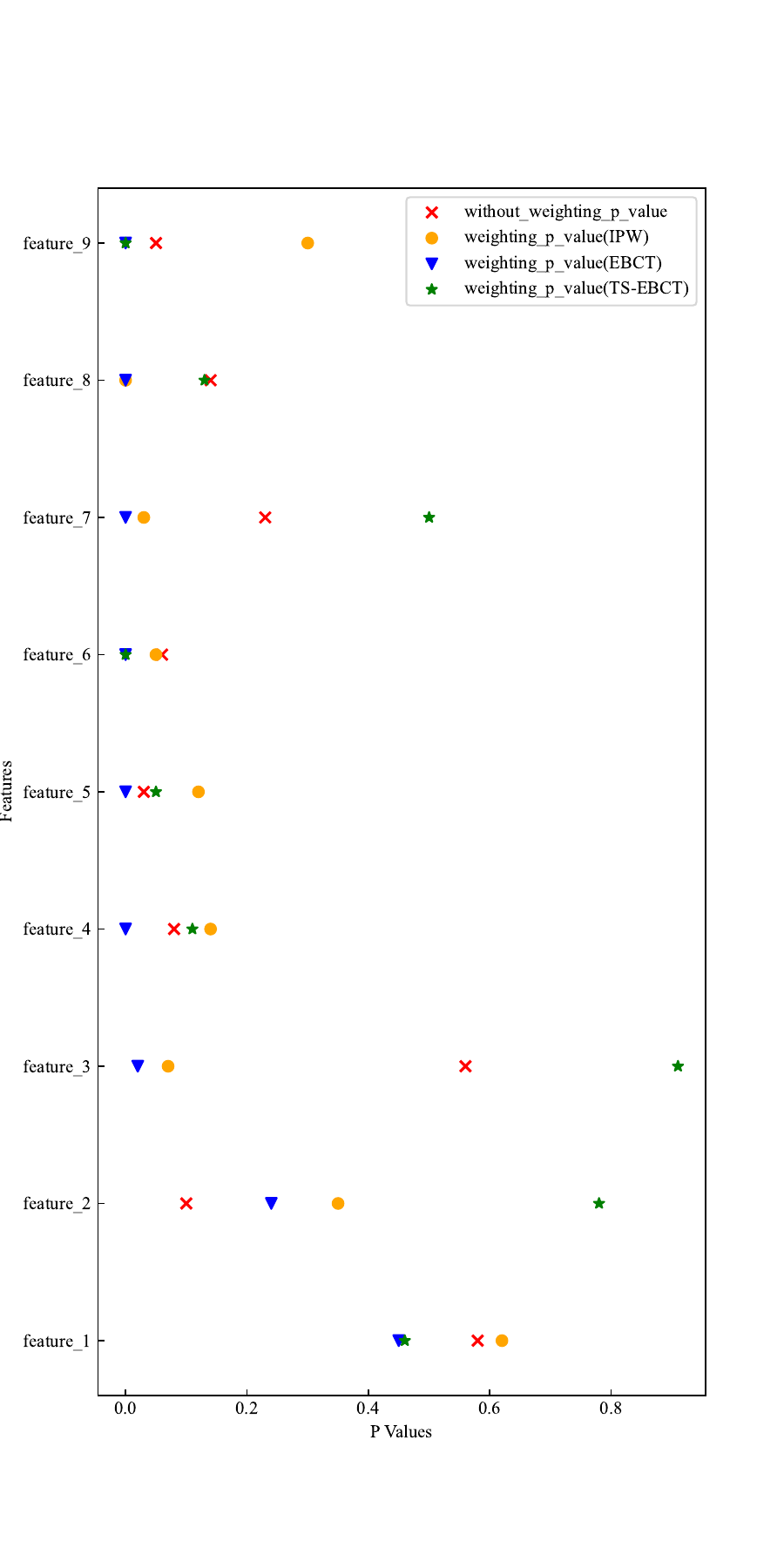}
    \caption{The feature elimination results of real dataset 2.}
    \label{fig:The feature elimination results of real dataset2}
\end{figure}

To verify the effectiveness of different re-weighting methods in eliminating the correlation between covariates and treatment variables, we conducted relevant model experiments on two simulated and two real datasets. To better illustrate the ability of re-weighting methods to eliminate bias, we selected some features with high correlation with the treatment variable from the 4 datasets for correlation visualization. The average absolute correlation value of all features can be found in the corresponding table, aiming to illustrate the overall change in the correlation between covariate features and treatment variables after re-weighting the samples. The following section will analyze the results of each dataset separately.
\paragraph{\textbf{Simulated dataset 1.}}
The data in Figure \ref{fig:The feature elimination results of simulation dataset1} reveals that TS-EBCT method outperforms in eradicating the correlation between these features and the treatment variable. Post sample re-weighting, the correlation with the treatment variable significantly diminishes compared to the unweighted method. While IPW and EBCT can also curtail the correlation between the feature and treatment variable to a certain degree, they simultaneously increase the correlation in some features. Considering the average correlation of all features, the TS-EBCT method reduces the feature correlation by 26.8\% in comparison to the unweighted method, whereas IPW and EBCT achieve a reduction of 11.3\% and 23.9\% respectively.
\paragraph{\textbf{Simulated dataset 2.}}
Notably showing in Figure \ref{fig:The feature elimination results of simulation dataset2}, among these features, TS-EBCT method demonstrated the most superior performance. Although the IPW and EBCT methods exhibited some instability across all features, they still yielded commendable results in effectively eliminating correlations overall. Specifically, when considering the average correlation across all features, IPW, EBCT, and TS-EBCT methods reduced the correlation by 10.3\%, 22.1\%, and 25\%, respectively, as compared to the unweighted method.
\paragraph{\textbf{Real dataset 1.}}
Figure. \ref{fig:The feature elimination results of real dataset1} shows that the TS-EBCT method reduces the correlation between the 9 features and the treatment variable after sample re-weighting, compared to the unweighted scenario. While IPW and EBCT methods manage to eliminate some correlation for most features, they are unstable for feature 1 and feature 7, even increasing correlation. Regarding the average absolute correlation of all features, all re-weighting methods lower feature correlation compared to non-reweighted methods. Entropy balance method outperforms IPW. After re-weighting with IPW, EBCT, and TS-EBCT, overall feature correlation decreases by 7.6\%, 18\%, and 44\% respectively. The data suggests that the TS-EBCT method excels in eliminating the bias between the covariate and treatment variable in temporal-spatial datasets, paving the way for more precise future causal effect learning.
\paragraph{\textbf{Real dataset 2.}}
As shown in Figure \ref{fig:The feature elimination results of real dataset2}, various re-weighting methods performed well in reducing bias for features 1, 2, and 3. However, for features 4, 5, and 7, IPW and EBCT re-weighting methods increased feature correlation instead of reducing it, which did not achieve the expected effect. TS-EBCT method performed relatively consistently, showing a significant decrease in correlation for features 4, 5, and 7. Compared to the unweighted approach, feature correlation decreased by 61\%, 33\%, and 50\% respectively. For features 6 and 8, all three re-weighting methods increased feature correlation. However, TS-EBCT has an increase of 80\% and 20\% respectively, while IPW and EBCT have much higher increases in correlation for these features. Overall, all three re-weighting methods were able to reduce feature correlation (as shown in Table \ref{table:Average feature correlation}), with IPW, EBCT, and TS-EBCT resulting in an average decrease of 13\%, 50\%, and 87.5\% compared to the unweighted approach.
\begin{table}[ht]
\caption{The average feature absolute correlation after applying weighting methods.}
\label{table:Average feature correlation}
\centering
\resizebox{\linewidth}{!}{
\begin{tabular}{|c|cccc|}
\hline
\multirow{2}{*}{dataset name} & \multicolumn{4}{c|}{average absolute correlation} \\ \cline{2-5} 
 & \multicolumn{1}{c|}{unweighted} & \multicolumn{1}{c|}{IPW} & \multicolumn{1}{c|}{EBCT} & TSEBCT(Ours) \\ \hline
simulated datasets 1 & \multicolumn{1}{c|}{0.071} & \multicolumn{1}{c|}{0.063} & \multicolumn{1}{c|}{0.054} & \textbf{0.052} \\ \hline
simulated datasets 2 & \multicolumn{1}{c|}{0.068} & \multicolumn{1}{c|}{0.061} & \multicolumn{1}{c|}{0.053} & \textbf{0.051} \\ \hline
real dataset 1 & \multicolumn{1}{c|}{0.13} & \multicolumn{1}{c|}{0.12} & \multicolumn{1}{c|}{0.11} & \textbf{0.09} \\ \hline
real dataset 2 & \multicolumn{1}{c|}{0.15} & \multicolumn{1}{c|}{0.13} & \multicolumn{1}{c|}{0.1} & \textbf{0.08} \\ \hline
\end{tabular}
}
\end{table}
\subsubsection{Performance on AUUC and AUC Metrics}
In this section, we mainly aim to verify the impact of the weights obtained by various re-weighting methods on causal effects and to determine whether they can enhance causal effects. The evaluation of causal effects is widely measured using the AUUC metric. A higher value indicates better performance in causal effect learning. We validated our approach on four datasets by using the weights obtained by IPW, EBCT, and TS-EBCT methods as the sample weights for the meta-learning model (e.g., S-learner) proposed by \cite{kunzel2019metalearners}. The base model of S-learner uses the popular machine learning model lightgbm. Alternatively, we can use the existing open-source package for causal inference, such as causalml \cite{chen2020causalml}, which provides a convenient implementation. Table \ref{table: auuc and auc} provides a summary of the experimental results, which will be analyzed in detail below.

Without sample weighting, the AUUC and AUC metrics of S-learner are 0.5796 and 0.9618 respectively on simulated dataset 1. Post sample weighting with IPW, both AUUC and AUC metrics remain largely unchanged. However, with EBCT and TS-EBCT weighting, AUUC metric increases by 0.0015 and 0.0048 respectively, with no significant AUC alteration. On simulated dataset 2, S-learner model's AUUC and AUC metrics, without sample weighting, are 0.5748 and 0.9619 respectively. After applying IPW, EBCT, and TS-EBCT weighting, the AUUC increases by 0.005, 0.009, and 0.01 respectively, while the AUC remains stable. On real dataset 1, the unweighted S-learner achieved an AUUC of 0.1592 and an AUC of 0.7878. When applying sample weighting via IPW, EBCT, and TS-EBCT, the AUUC metric improvements for the S-learner were 0.25\%, 0.87\%, and 1.7\%, respectively. Notably, these weighting methods had minimal impact on the AUC of the S-learner. On real dataset 2, the S-learner achieved an AUUC of 0.049 and an AUC of 0.748 without sample weighting. The TS-EBCT method demonstrated superior AUUC improvement compared to EBCT when re-weighting the S-learner. Both entropy balance methods (TS-EBCT and EBCT) outperformed IPW in sample weighting, indicating their effectiveness in mitigating the influence of confounding variables and enhancing causal effect estimation. The AUUC metric improvements for IPW, EBCT, and TS-EBCT were 7.5\%, 15\%, and 27\%, respectively, compared to the unweighted scenario.
\begin{table}[ht]
\caption{The enhancement effect of different reweighting methods on AUUC and AUC metrics was separately validated on four datasets.}
\label{table: auuc and auc}
\centering
\resizebox{\linewidth}{!}{
\begin{tabular}{|c|c|cccc|}
\hline
\multirow{2}{*}{datasets}    & \multirow{2}{*}{metrics} & \multicolumn{4}{c|}{compared methods}                                                                              \\ \cline{3-6} 
                        &                    & \multicolumn{1}{c|}{unweighted}    & \multicolumn{1}{c|}{IPW}    & \multicolumn{1}{c|}{EBCT}   & TS-EBCT(Ours) \\ \hline
\multirow{2}{*}{simulated dataset 1} & AUC                & \multicolumn{1}{c|}{0.9618} & \multicolumn{1}{c|}{0.9614} & \multicolumn{1}{c|}{0.9612} & 0.9611       \\ \cline{2-6} 
                        & AUUC               & \multicolumn{1}{c|}{0.5796} & \multicolumn{1}{c|}{0.5797} & \multicolumn{1}{c|}{0.5811} & \textbf{0.5844}       \\ \hline
\multirow{2}{*}{simulated dataset 2} & AUC                & \multicolumn{1}{c|}{0.9619} & \multicolumn{1}{c|}{0.9617} & \multicolumn{1}{c|}{0.9617} & 0.9614       \\ \cline{2-6} 
                        & AUUC               & \multicolumn{1}{c|}{0.5748} & \multicolumn{1}{c|}{0.5798} & \multicolumn{1}{c|}{0.5839} & \textbf{0.585}       \\ \hline
\multirow{2}{*}{real dataset 1} & AUC                & \multicolumn{1}{c|}{0.7878} & \multicolumn{1}{c|}{0.7875} & \multicolumn{1}{c|}{0.7873} & 0.7874       \\ \cline{2-6} 
                        & AUUC               & \multicolumn{1}{c|}{0.1592} & \multicolumn{1}{c|}{0.1596} & \multicolumn{1}{c|}{0.1606} & \textbf{0.162}        \\ \hline
\multirow{2}{*}{real dataset 2} & AUC                & \multicolumn{1}{c|}{0.748}  & \multicolumn{1}{c|}{0.747}  & \multicolumn{1}{c|}{0.746}  & 0.746        \\ \cline{2-6} 
                        & AUUC               & \multicolumn{1}{c|}{0.049}  & \multicolumn{1}{c|}{0.053}  & \multicolumn{1}{c|}{0.058}  & \textbf{0.068}        \\ \hline
\end{tabular}
}
\end{table}
\section{Conclusion}
The entropy balancing method has been widely used in recent years, and the base weights it generates bring new possibilities for further improving the accuracy of causal inference. In this paper, we proposed TS-EBCT method, which can further eliminate the influence of confounding variables in the temporal-spatial dimension, thereby more accurately learning the causal effects of the temporal-spatial dimension, improving the accuracy of the causal inference model in the logistics industry, and bringing higher returns to the company's subsidy and pricing business. We verified the effectiveness and convergence of the TS-EBCT method on two simulated and two real datasets, respectively. For example, on the real dataset 2, the TS-EBCT method reduces the average absolute correlation index by about 46\% compared to the unweighted method, 38.4\% compared to the IPW method, and 20\% compared to the EBCT method. In addition, the base weights obtained by the TS-EBCT method also play a significant role in improving the accuracy of causal effects. For example, on the real dataset 2, the basic weights solved by the TS-EBCT method improve the AUUC index of the S-Learner samples by about 27\% after weighting. Furthermore, in the future, we will continue to explore how to solve and accelerate algorithms under large-scale data volumes and high-dimensional feature spaces, and try to combine deep learning methods for end-to-end optimization work.
\bibliography{references}
\bibliographystyle{icml2022}
\end{document}